\documentclass[floatfix,aps,tightenlines,prd,nofootinbib,superscriptaddress,twocolumn]{revtex4}

\usepackage{epsfig}
\usepackage{graphics}
\usepackage{amsmath}

\newcommand{\sA}{\textrm{{\scriptsize A}}}
\newcommand{\tA}{\textrm{{\tiny A}}}
\newcommand{\sB}{\textrm{{\scriptsize B}}}
\newcommand{\sC}{\textrm{{\scriptsize C}}}
\newcommand{\eup}[2]{\ensuremath{\mathrm{e}^{#1}_{\cdot #2}}}
\newcommand{\edn}[2]{\ensuremath{\mathrm{e}^{\cdot #1}_{#2}}}
\newcommand{\ecap}[1]{\ensuremath{\mathbf{\hat{e}}_{#1}}}
\newcommand{\thetacap}[1]{\ensuremath{\mathbf{\hat{\theta}}^{#1}}}
\newcommand{\K}[3]{\ensuremath{ \mathrm{K}^{#1}_{\cdot \, #2 #3}}}
\newcommand{\T}[3]{\ensuremath{ \mathrm{T}^{#1}_{\cdot \, #2 #3}}}
\newcommand{\W}[3]{\ensuremath{ \mathrm{\omega}^{#1}_{\cdot \, #2 #3}}}
\newcommand{\Ti}[3]{\ensuremath{\mathrm{T}^{\cdot \, #1}_{#2\, \cdot #3}}}
\newcommand{\Chris}[3]{\ensuremath{\Gamma^{#1}_{\cdot \, #2 #3}}}
\newcommand{\Christilde}[3]{\ensuremath{ \tilde{\Gamma}^{#1}_{\cdot \, #2 #3}}}
\newcommand{\Chrishat}[3]{\ensuremath{ \hat{\Gamma}^{#1}_{\cdot \, #2 #3}}}

\newcommand{\A}{\ensuremath{\mathrm{A}}}
\newcommand{\J}{\ensuremath{\mathrm{J}}}

\newcommand{\h}{\ensuremath{\mathrm{H}}}
\newcommand{\B}{\ensuremath{\mathrm{B}}}

\newcommand{\5}{\ensuremath{\dot{5}}}
\newcommand{\nablatilde}{\tilde{\nabla}}

\begin{document}
\title{Kaluza-Klein Theory with Torsion confined to the Extra-dimension}
\author{Karthik H. Shankar
and Kameshwar C. Wali
}
\affiliation{Department of Physics, Syracuse University, Syracuse, NY-13244, USA}

\begin{abstract}
Here we consider a variant of the 5 dimensional Kaluza-Klein theory within the framework of Einstein-Cartan formalism that includes torsion. By imposing a set of constraints on torsion and Ricci rotation coefficients, we show that the torsion components are completely expressed in terms of the metric. Moreover, the Ricci tensor in 5D corresponds exactly to what one would obtain from torsion-free general relativity on a 4D hypersurface. The contributions of the scalar and vector fields of the standard K-K theory  to the Ricci tensor and the affine connections are completely nullified by the  contributions from torsion. As a consequence, geodesic motions do not distinguish the torsion free 4D space-time from a hypersurface of  5D space-time with torsion satisfying the constraints.  Since torsion is not an independent dynamical variable in this formalism, the modified Einstein equations are different from those in the general Einstein-Cartan theory. This leads to important cosmological consequences such as the emergence of cosmic acceleration.
\end{abstract}

\maketitle{}

In any attempt to link fundamental matter fields with intrinsic spin to gravity, it becomes necessary to extend the Riemannian space-time to include \emph{torsion}, defined to be the antisymmetric part of affine connection. The resulting theory of gravity, known as the Einstein-Cartan theory\cite{Hehl} treats the metric and torsion as two independent geometrical characteristics of space-time \cite{Shapiro}.

Historically, beginning with the  Kaluza-Klein (KK)  theory, there has been a great interest in introducing extra dimensions of space-time to unify gravity with elementary particle interactions. In the KK theory, the scalar and vector fields, which are the extra dimensional components of the metric tensor contribute to the affine connection and the Ricci tensor and hence modify their values from the corresponding values in 4D space-time\cite{Wesson}. The contribution of these fields  to the Einstein tensor are normally interpreted as gravity induced matter.

\begin{figure}
\begin{center}
\includegraphics[scale=0.30]{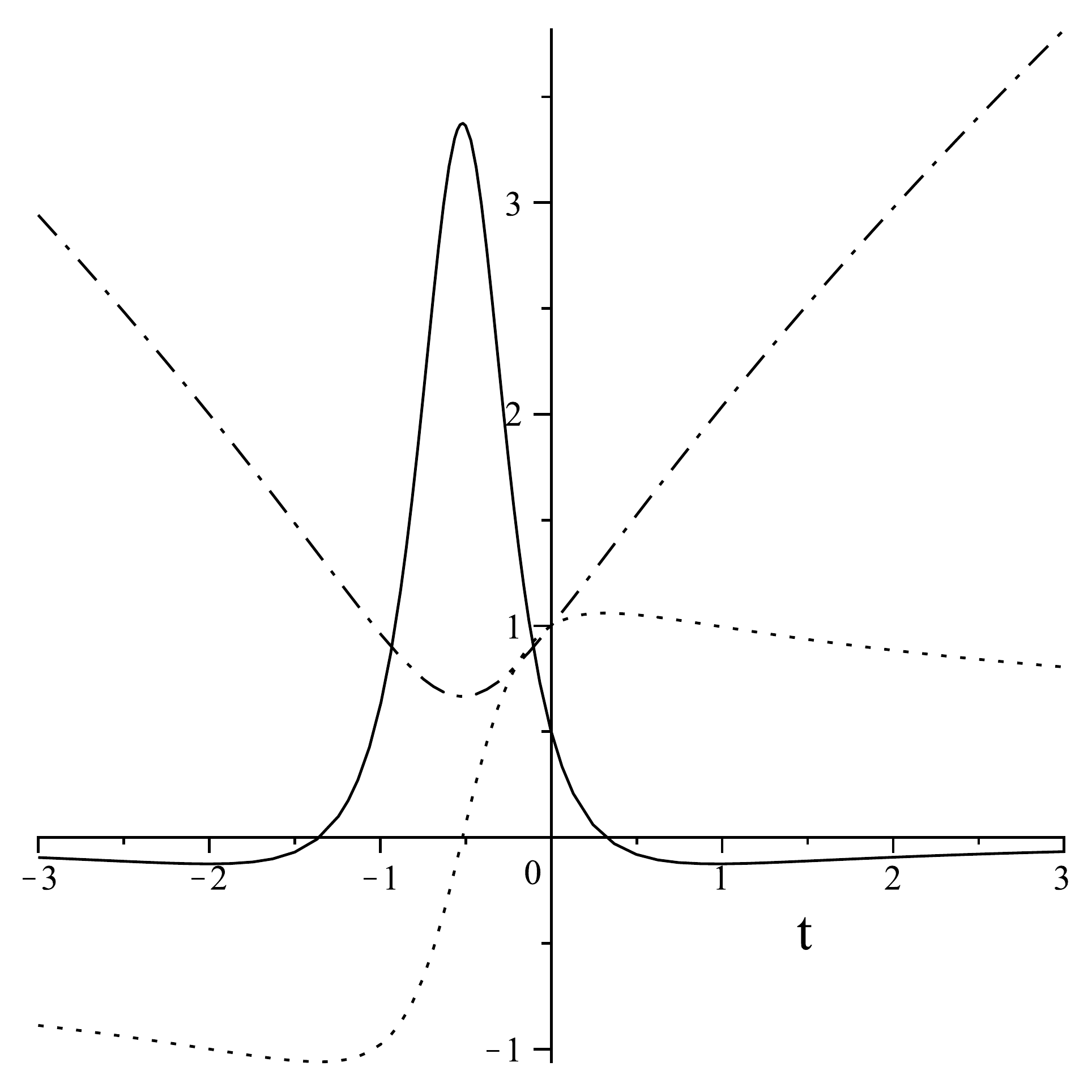}
\caption{The dashed-dot curve denotes $a(t)$, the dotted curve denotes $\dot{a}(t)$, and the solid curve denotes 
$\overset{..}{a}(t)$. The time axis is in units of $\h_{o}^{-1}$. }
\label{FRW}
\end{center}
\end{figure}

In this work, we incorporate torsion into 5D KK theory\cite{Kalinowski}. The inclusion of torsion introduces free parameters in the affine connection and the Ricci tensor in addition to  the contributions from extra dimensional metric components that occur in the torsion free KK theory. In this paper we impose a minimal set of conditions so as to restrict torsion to purely extra dimensional components and determine all its components in terms of  the metric. Interestingly, the imposed conditions lead to a complete cancellation between the modifications induced by the extra dimensional metric components and the contributions from the torsion. Thus the Ricci tensor in 5D space-time in the resulting formalism is exactly the Ricci tensor in a torsion free 4D space-time. 
In the second part of the paper, we apply the action principle to derive the equations of motion of this formalism. The modified Einstein equations thus derived are finally applied to Robertson-Walker cosmology that leads to a novel expansion history for the universe (see figure \ref{FRW}).  

To  describe the constraints to be imposed, we start, for the sake of completeness, with a brief overview on the relationship between tetrads, torsion and affine connection.

$\\$

\textbf{Setting up the framework:}
Since the constraints we propose to impose can be more simply
stated with reference to a locally flat inertial system (allowed
by the equivalence principle), we proceed to define and collect
together from reference \cite{Nakaharabook}, the relevant standard relations we
need in both the coordinate and the inertial frames. Let ($i,
j,k,...$) and ($\sA,\sB,...$) denote coordinate and inertial frame
indices  respectively and $\ecap{i}=\partial_{i}$ and
$\thetacap{i}=dx^{i}$ be the basis of the tangent and dual spaces
at each point in space-time. We define the corresponding inertial
basis to be $\ecap{\sA}=\eup{i}{\sA} \ecap{i}$ and
$\thetacap{\sA}=\edn{\sA}{i}\thetacap{i}$, where the vielbeins
$\eup{i}{\sA}$ and $\edn{\sB}{j}$  satisfy the orthonormality
conditions,
\begin{equation}
\edn{\sA}{i} \eup{j}{\sA}={\delta}^{j}_{i}; \qquad
\edn{\sA}{i}\eup{i}{\sB}={\delta}^{\sA}_{\sB}.
\label{eq:orthonormality}
\end{equation}

By definition, the metric in the inertial frame is Minkowskian
$\eta_{AB}$, and the metric tensor in the coordinate system is
$\mathbf{g}_{i j}=\edn{\sA}{i} \edn{\sB}{j} \eta_{AB}$ and
$\mathbf{g}^{i j}=\eup{i}{\sA} \eup{j}{\sB} \eta^{AB}$. The
covariant derivative operator ($\tilde{\nabla}$) can be defined in terms of the
coordinate basis, or equivalently in terms of the inertial frame
basis,
\begin{equation}
\nablatilde_{\ecap{i}}\ecap{j}=\Christilde{k}{i}{j}\ecap{k}, \qquad
\nablatilde_{\ecap{\tA}}\ecap{\sB}=\W{\sC}{\sA}{\sB}\ecap{\sC},
\end{equation}
where $\Christilde{i}{j}{k}$ and $\W{\sA}{\sB}{\sC}$ are the
affine and the Ricci rotation coefficients respectively. The
relationship between these two quantities follows from the
transformation laws between the coordinate frame(\ecap{i}) and
inertial frame(\ecap{\sA}),
\begin{equation}
\W{\sA}{\sB}{\sC}=\eup{i}{\sB} ( \nablatilde_{\ecap{i}} \eup{j}{\sC} )
\edn{\sA}{j}.
\label{def:omega}
\end{equation}

The affine connection by itself is not a tensor,
but its antisymmetric part, the torsion, is a tensor. 
\footnote{Throughout this paper, square brackets enclosing indices denote the conventional anti-symmetrization, while the regular brackets enclosing the indices denotes symmetrization.}
\begin{equation}
\T{i}{j}{k}=\Christilde{i}{j}{k}-\Christilde{i}{k}{j} = 2 \Christilde{i}{[j}{k]}
\end{equation}
Again, using the transformation laws between the coordinate and
the inertial frames, we have
\begin{equation}
\T{i}{j}{k} = \edn{\sB}{j}\edn{\sC}{k}
\eup{i}{\sA}\T{\sA}{\sB}{\sC}. \label{torsion_transform}
\end{equation}

Furthermore, with the standard assumption of metric compatibility,
namely $\nabla_{\ecap{i}} g_{jk}=0$, we obtain
\begin{equation}
\Christilde{i}{j}{k} = \Chrishat{i}{j}{k} + \K{i}{j}{k},
\label{chris1}
\end{equation}
where $\Chrishat{i}{j}{k}$ is the Christoffel connection,  
\begin{equation}
\Chrishat{i}{j}{k}= \Big\{ {}_{j} {}^{i} {}_{k} \Big\}
=\frac{1}{2}\mathbf{g}^{im}[{\partial_{j}\mathbf{g}_{km}+\partial_{k}\mathbf{g}_{jm}-\partial_{m}\mathbf{g}_{jk}}] ,
\label{def:Chris}
\end{equation}
and $\K{i}{j}{k}$ is the contorsion tensor.
\begin{equation}
\K{i}{j}{k}= \frac{1}{2} \left[ \T{i}{j}{k} + \Ti{i}{j}{k} + \Ti{i}{k}{j} \right] .
\label{def:contorsion}
\end{equation}

\textbf{Constructing the 5D space-time:}
We shall now focus on applying the above formalism to a five
dimensional space-time. Consider a foliation of the 5D space-time
in terms of a family of 4D hypersurfaces, which are parametrized
by the coordinate system $\{x^{\mu} \}$, where $(\mu, \nu,...)$
denote the coordinate indices  on these hypersurfaces. Let $x^{5}$
denote the parametrization of the family, and $5$ denote the
corresponding coordinate index. The hypersurface coordinates
$\{x^{\mu}\}$ together with $x^{5}$ will then span the entire
5D space-time. Let the metric and its inverse on each of the
hypersurface be $\mathrm{g}_{\mu \nu}$ and $\mathrm{g}^{\mu \nu}$
respectively, which can in principle depend on the $x^{5}$
coordinate. Let $\{\eup{\mu}{a},\edn{a}{\mu}\}$ denote the tetrad
system on these hypersurfaces satisfying the orthonormality
relations $\eup{\mu}{a} \edn{b}{\mu}=\mathbf{\delta}^{a}_{b}$ and
$\eup{\mu}{a}\eup{a}{\nu}=\mathbf{\delta}^{\mu}_{\nu}$. Here
$(a,b,...)$ denote the tetrad indices on these 4D hypersurfaces.
The metric on the hypersurface
 is then given by $\mathrm{g}_{\mu \nu}=\edn{a}{\mu}
\edn{b}{\nu} \eta_{ab}$ and  its inverse $\mathrm{g}^{\mu
\nu}=\eup{\mu}{a} \eup{\nu}{b} \eta^{ab}$.

We will now construct  the vielbiens in the 5D space-time by
extending the tetrad system on the 4D hypersurfaces. We take the
components of the 5D vielbeins to be $\eup{i}{\sA}=(\eup{\mu}{a},
\eup{\mu}{\5}, \eup{5}{a}, \eup{5}{\5})$ and $\edn{\sA}{i}= 
(\edn{a}{\mu}, \edn{a}{5}, \edn{\5}{\mu},\edn{\5}{5} )$. 
The index $\5$ corresponds to the fifth dimension of the inertial frame.
The orthonormality relations in 5D (eq. \ref{eq:orthonormality})
immediately leads to
\footnote{ There exists another class of vielbeins that satisfy the orthonormality relations.
But they are essentially related to eq. \ref{eq:vielbeins} by gauge. Here, we choose to work
with the vielbeins given by  eq. \ref{eq:vielbeins} to make the formalism readily  comparable
to the existing Kaluza-Klein literature.}
\begin{eqnarray}
 \eup{\mu}{\5}=0, \,\, \eup{5}{a}=-\eup{\mu}{a}\mathrm{\A}_{\mu}, \, \, \eup{5}{\5} = \Phi^{-1},  \nonumber \\
 \edn{a}{5}=0, \,\, \edn{\5}{\mu}=\A_{\mu} \Phi, \,\, \edn{\5}{5}= \Phi.
\label{eq:vielbeins}
\end{eqnarray}

The metric in the 5D is then given by $\mathbf{g}_{i j}=\edn{\sA}{i}
\edn{\sB}{j} \eta_{AB}$ and $\mathbf{g}^{i j}=\eup{i}{\sA} \eup{j}{\sB}
\eta^{AB}$.

\begin{eqnarray}
\mathbf{g}_{\mu \nu} &=& \mathrm{g}_{\mu \nu}+\epsilon \A_{\mu}\A_{\nu}\Phi^{2},\,
\mathbf{g}_{\mu 5} = \epsilon \A_{\mu}\Phi^{2}, \,
\mathbf{g}_{5 5}=\epsilon \Phi^{2}, \nonumber \\
\mathbf{g}^{\mu \nu} &=& \mathrm{g}^{\mu \nu}, \,
 \mathbf{g}^{\mu 5}=-\A^{\mu}, \,
\mathbf{g}^{5 5} = \A_{\lambda} \A^{\lambda} + \epsilon \Phi^{-2} .
\label{eq:5Dmetric}
\end{eqnarray}

The raising and lowering of indices on the
vector field is done w.r.t the 4D metric
$\mathrm{g}_{\mu \nu}$. The parameter $\epsilon= \pm 1$ denotes  whether the extra dimension is space-like or time-like. Note that the induced metric on the hypersurfaces (induced by the 5D geometry), $\mathrm{g}_{\mu \nu}+\epsilon \A_{\mu}\A_{\nu}\Phi^{2}$ is different from the 4D metric  $\mathrm{g}_{\mu \nu}$ on them, but quite evidently related by a gauge transformation. 

We will now impose a set of constraints on torsion and the Ricci
rotation coefficients consistent with Cartan's structure
equation \cite{Nakaharabook} that relates torsion and connection
coefficients. With a minimal modification of the
standard general relativity in mind, we chose the set of
constraints \cite{Viet} so that the torsion components tangential to the 4D hypersurface
vanish (see also \cite{Bohmer}), while leaving non-vanishing torsion components completely
determined in terms of the metric components of the 5D space-time.
In the absence of torsion, the metric compatibility condition serves to evaluate the connection coefficients (Christoffel symbols, eq. \ref{def:Chris}) in terms of the metric components. In the presence of torsion, the imposed constraints can be viewed as an extension to the metric compatibility condition in the spirit that they together serve to determine the connection coefficients in terms of the metric components. Furthermore, the imposed constraints are covariant conditions on the 4D hypersurface. That is, the conditions are form invariant with respect to within-hypersurface coordinate transformations (that do not mix $\mu,\nu$ with $5$).  

\[
\textbf{Condition 1} :  \T{a}{\sB}{\sC}=0
\]

Using eq. \ref{torsion_transform} and noting that
$\eup{\mu}{\5}=0$, we find $\T{\mu}{i}{k}=0$. This implies that
the only nonzero components of torsion are \T{5}{i}{k}. To
determine these, we impose the following condition on the Ricci
rotation coefficients,
\[
\textbf{Condition 2} :  \W{\5}{\sB}{\sC}=0
\]
This condition along with metric compatibility implies
$\W{\sA}{\sB}{\5}=0$. We can now use eq. \ref{def:omega} to write,
\begin{equation}
\W{\sA}{\sB}{\5}=
\eup{i}{\sB} ( \nablatilde_{\ecap{i}} \eup{j}{\5}) \edn{\sA}{j}=
\eup{i}{\sB} (
\partial_{i} \eup{j}{\5} + \Christilde{j}{i}{k} \eup{k}{\5} )
\edn{\sA}{j} =0
\end{equation}
Since $\eup{\mu}{\5}= \eup{a}{5}=0$ , the above equation implies,
\begin{equation}
\Christilde{\mu}{i}{5}=0, \qquad
\Christilde{5}{i}{5} =
-\edn{\5}{5} \partial_{i} \eup{5}{\5}
\end{equation}
Using the above equations along with eq.\ref{chris1}, we can
express the contorsion in terms of the Christoffel symbols
\Chrishat{}{}{}and the vielbeins,
\begin{eqnarray}
\K{\mu}{i}{5} &=& -\Chrishat{\mu}{i}{5},  \nonumber \\
\K{5}{i}{5}&=&-\left( \Chrishat{5}{i}{5} + \edn{\5}{5}
\partial_{i} \eup{5}{\5} \right)
 =- \Chrishat{5}{i}{5} + \mathrm{J}_{i},
 \label{c2}
\end{eqnarray}
where $\mathrm{J}_{i} \equiv  \Phi^{-1} \partial_{i}\Phi$.

Note that the above equations give only a subset of the components \K{i}{j}{k}, but along with eq. \ref{def:contorsion}, they are sufficient to determine all the components of the torsion. The torsion thus obtained can in turn be substituted in eq. \ref{def:contorsion} to determine the remaining components of contorsion. The components of torsion are
\begin{eqnarray}
\T{\mu}{i}{j}&=&\T{5}{5}{5}= 0, \nonumber \\
\T{5}{\mu}{\nu} &=& 2 \partial_{[\mu} \A_{\nu]} +2 \J_{[\mu} \A_{\nu]},  \nonumber \\
\T{5}{\mu}{5} &=& \J_{\mu} - \partial_{5} \A_{\mu} -\A_{\mu} \J_{5}, 
\label{torsions}
\end{eqnarray}

In addition to yielding the non-vanishing components of torsion in
terms of the metric components, the solution to eq. \ref{c2} also
yields the following condition on the  metric on the 4D
hypersurfaces.
\begin{equation}
\partial_{5} \mathrm{g}_{\mu \nu}=0.
\end{equation}


This implies all the hypersurfaces in the foliating family have the same
4D metric. To place things in perspective, we observe that in
the standard Kaluza-Klein theory, the assumption of cylindrical
condition makes all the quantities, namely $\mathrm{g}_{\mu \nu}$,
$\A_{\mu}$ and $\Phi$ independent of $x^{5}$. Whereas in our
formulation, the constraints automatically imply $\mathrm{g}_{\mu
\nu}$ is independent of $x^{5}$, while $\A_{\mu}$ and  $\Phi$
could still depend on $x^{5}$.

With all the \K{i}{j}{k} determined from eqns. \ref{torsions} and \ref{def:contorsion}, we now use eq.
\ref{chris1} to calculate all the affine connection coefficients.
\begin{eqnarray}
\Christilde{\lambda}{5}{5}&=&\Christilde{\lambda}{\nu}{5}=\Christilde{\lambda}{5}{\nu}=0, \nonumber \\
\Christilde{5}{\mu}{\nu}&=&\nabla_{\mu}\A_{\nu}+ \J_{\mu} \A_{\nu}, \nonumber \\
\Christilde{5}{5}{\mu}&=& \partial_{5} \A_{\mu} + \J_{5} \A_{\mu}, \nonumber \\
\Christilde{5}{\mu}{5} &=& \J_{\mu}, \,\,\, \Christilde{5}{5}{5} =
\J_{5}, \,\,\,
\Christilde{\lambda}{\mu}{\nu}=\Chris{\lambda}{\mu}{\nu}.
\label{connections}
\end{eqnarray}

Here \Chris{\lambda}{\mu}{\nu} corresponds to the Christoffel
symbols (analogous to eq. \ref{def:Chris}) obtained from
torsion free 4D space-time with metric  $\mathrm{g}_{\mu \nu}$. Note that
the components of 5D Christoffel symbols along the hypersurface
coordinates is not equal to the Christoffel symbols calculated on
4D spacetime, that is, $\Chrishat{\lambda}{\mu}{\nu} \neq
\Chris{\lambda}{\mu}{\nu} $. The symbol $\nabla_{\mu}$ corresponds
to the covariant derivative operator on the torsion-free 4D
space-time, where the Christoffel symbols are  exactly the
affine connection coefficients.

Substituting the above connection coefficients in the Ricci tensor
defined by
\begin{equation}
\tilde{R}_{ik}=\partial_{k}\Christilde{j}{j}{i} - \partial_{j} \Christilde{j}{k}{i}
+\Christilde{j}{k}{m}\Christilde{m}{j}{i}- \Christilde{j}{j}{m}\Christilde{m}{k}{i},
\end{equation}
we find
\begin{equation}
\tilde{R}_{\mu \nu}= R_{\mu \nu}, \,\, \tilde{R}_{\mu 5} = \tilde{R}_{5 \mu} = \tilde{R}_{5 5}=0.
\end{equation}
Here $R_{\mu \nu}$  represents the Ricci tensor
on the torsion-free 4D space-time. It also follows that the 5D Ricci
scalar is exactly the same as the Ricci scalar in the torsion free
4D space time,  that is $\tilde{R} =R$. 
We also note, in general, the presence of torsion
makes the Ricci tensor non-symmetric, but the constraints we have
imposed on the torsion leaves the Ricci tensor symmetric. 

It is straightforward to see that the formalism and the results obtained thus far are not specific to 
4 and 5 dimensions, they can be generalized to any
arbitrary D and D+1  dimensions.
We shall now consider some implications of the formalism with respect to geodesic equations
and solutions to Einstein equations.

$\\$

\textbf{Geodesic Equations:} The 5D geodesic equations split into
\begin{eqnarray}
&\overset{..}{x}^{5}& + \Christilde{5}{\mu}{\nu}  \dot{x}^{\mu}
\dot{x}^{\nu} + \left( \Christilde{5}{\mu}{5} +
\Christilde{5}{5}{\mu}  \right)  \dot{x}^{\mu} \dot{x}^{5}
+ \Christilde{5}{5}{5} \left( \dot{x}^{5} \right)^{2}=0, \nonumber \\
&\overset{..}{x}^{\lambda}& +\Chris{\lambda}{\mu}{\nu}
\dot{x}^{\mu} \dot{x}^{\nu} =0
\label{eq:geodesic}
\end{eqnarray}
 We note that the components of the geodesic equations along the hypersurface
are exactly the same as the geodesic equations in the torsion free 4D space-time.  This is in contrast with the 
conventional Kaluza Klein theory where the 4D geodesic equations are modified by the presence of 
the fields $\A_{\mu}$ and $\Phi$. In this formalism, the presence of torsion completely nullifies the effect of 
these fields in the 4D geodesic equations.
Furthermore, it is worth noting that the geodesic of a particle can be confined to a 4D hypersurface by requiring 
$\dot{x}^{5}=0$. From eq. \ref{eq:geodesic}, we see that this requires the additional condition, 
\begin{equation}
\Christilde{5}{\mu}{\nu}=\nabla_{\mu}\A_{\nu}+ \J_{\mu} \A_{\nu}=0. 
\end{equation}
If this condition can be satisfied, then there will be no observable difference, as far as a test particle is concerned, whether we live in a torsion free 4D space-time or on a hypersurface within the 5D space-time with torsion. However, it is apparent that this condition is  a strong constraint requiring the vector and scalar fields satisfy the above equation for a given 4D metric. It is conceivable that for some 4D  metrics, no choice of vector and scalar fields would satisfy the above constraint. In such cases, the particle would be free to move in the $x^{5}$ direction unless constrained by an external force or if the fifth co-ordinate is compact and small as usually assumed in most  adaptations of the Kaluza-Klein theory.


$\\$

\textbf{Einstein's equations:}  
To obtain the equations of motion, we need to vary the action with respect to the independent dynamical fields of the theory. In general, if one includes torsion in the theory, torsion and metric are two independent dynamical fields and consequently one needs to vary the action with respect to both these variables [1, 4]. In our approach, with torsion determined in terms of the metric, metric components are the only dynamical variables. Taking the action to be $S= \int \tilde{R} \sqrt{-\mathbf{g}} \, d^{5}x $, its variation yields
\footnote{ We are using geometric units where $G=c=1$. We also note that the action in general will include a matter Lagrangian term which would also contribute to the variation.}   
\begin{equation}
 \int \left[   \tilde{R} \, \delta \sqrt{-\mathbf{g}} 
+   \tilde{R}_{i k}  \sqrt{-\mathbf{g}} \, \delta \mathbf{g}^{ik}  
+ \mathbf{g}^{ik} \sqrt{-\mathbf{g}} \, \delta \tilde{R}_{i k}  \right] d^{5}x
\end{equation}
One can set this variation to zero to obtain the modified Einstein equations. The first two terms give rise to the usual symmetric Einstein tensor. In the absence of torsion, the third term becomes a boundary integral that contributes nothing to the equation. While in the presence of torsion, the third term can be shown to contribute   
\begin{equation}
\int  \left[  \T{m}{j}{m} \mathbf{g}^{ik}  \delta \Christilde{j}{k}{i}  
- \T{m}{k}{m} \mathbf{g}^{ik} \delta \Christilde{j}{j}{i}  
+  \T{m}{k}{j}  \mathbf{g}^{ik} \delta \Christilde{j}{m}{i} \right] \sqrt{-\mathbf{g}}  \, d^{5}x \nonumber
\end{equation}
Treating the variation in the torsion and metric components independent, one can obtain the Einstein-Cartan equations derived by Hehl et.al \cite{Hehl}. However, by expressing $\delta \Christilde{}{}{}$ purely in terms of variation of the 5D metric components (from eq. \ref{connections}), it turns out that the above expression can be simplified to 
\begin{equation}
\int \h_{\mu \nu} \delta g^{\mu \nu} \, \sqrt{-\mathbf{g}} \, d^{5}x
\label{extraterm}
\end{equation}   
where 
\begin{eqnarray}
\h_{\mu \nu} &\equiv& \nabla_{( \mu} \B_{\nu) } - (\nabla \cdot \B ) g_{\mu \nu}  + \J_{( \mu} \B_{\nu )} - (\J \cdot \B) g_{\mu \nu},  \nonumber \\
\B_{\mu} & \equiv&\T{5}{\mu}{5} = \J_{\mu} - \partial_{5} \A_{\mu} -\A_{\mu} \J_{5},
\end{eqnarray}
are interpreted as tensors in torsion free 4D space. 
Since eq. \ref{extraterm} is the only additional term that contributes to the variation of action, $\h_{\mu \nu}$ is essentially  the modification to the standard torsion free Einstein tensor. The modified Einstein equations then turns out to be  
\begin{eqnarray}
R_{\mu}^{\,\, \nu} -\frac{1}{2} R \delta^{\,\, \nu}_{\mu} + \h^{\,\, \nu}_{\mu} &=& \Sigma_{\mu}^{\,\, \nu}
\label{EE1} \\
-\A^{\alpha} R_{\mu \alpha} - \A^{\alpha} \h_{\mu \alpha} &=& \Sigma_{\mu}^{\,\, 5}
\label{EE2} \\
 -\frac{1}{2} R &=& \Sigma_{5}^{\,\, 5} 
 \label{EE3}
\end{eqnarray}
Here $\Sigma$ is the stress tensor that one would obtain when matter fields are included in the Lagrangian prior to variation of action. Its 4D components will be the observed stress energy tensor and can be identified with the stress energy tensor of the usual 4D torsion free general relativity. Its conservation immediately yields  
\begin{equation}
\nabla_{\nu}  \Sigma^{\,\, \nu}_{\mu} =0  \Longrightarrow \nabla_{\nu}  \h^{\,\, \nu}_{\mu} =0. 
\label{conservation}
\end{equation}
In general, the components $\Sigma_{\mu}^{\,\, 5}$ and $\Sigma_{5}^{\,\, 5}$ cannot be obtained from observations on the 4D hypersurface. Hence, without an explicit 5D matter Lagrangian specifying these components, it is not possible to solve equations \ref{EE2} and \ref{EE3}. Thus, we are left with just equations \ref{EE1} and \ref{conservation} to solve for the metric components $g_{\mu \nu}$, $\A_{\mu}$ and $\Phi$.

$\\$
 
\textbf{Robertson-Walker cosmology:}
To illustrate an application of the formalism, let us consider 
spatially flat homogenous and isotropic universe with metric
\begin{equation}
\mathrm{ds}^2 = -dt^2 +a^{2}(t) \left( dx^{2} + dy^{2} + dz^{2}
\right).
\end{equation}

The assumption of homogeneity and isotropy of the 4D geometry requires that $\A_{\mu} =(\A_{t}, 0,0,0)$ and  $\A_{t}$ and $\Phi$ are functions of $t$ and not the spatial coordinates. To further simplify, we shall also assume that the fields do not depend on $x^{5}$. 
From eq. \ref{torsions}, these constraints imply that $\B_{\mu}=\J_{\mu}$ and the only non vanishing component of $\J_{\mu}$ is $\J_{t}$, and of $\A_{\mu}$ is $\A_{t}$.
Applying the conservation equation (eq. \ref{conservation}) yields 
\begin{eqnarray} 
(i) \, \dot{ \Phi}= 0, \J_{t}=0,
 \,\, \mathrm{or} \qquad  
(ii) \, \Phi= \dot{a}(t), \J_{t}=\overset{..}{a}/\dot{a}. \\ \nonumber
\end{eqnarray}
 
Case $(i)$ yields $\h_{\mu \nu}=0$, and eq. \ref{EE1} yields the usual Friedman equation along with matter conservation.
\[
 \left( \dot{a}/a \right)^{2}  = \frac{8 \pi}{3} \rho.
\]
Case $(ii)$ yields non vanishing  $\h_{\mu \nu}$, which when applied to eq. \ref{EE1} gives the modified Friedman equation
\begin{equation} 
 \left( \dot{a}/a \right)^{2} +  \left( \overset{..}{a}/a \right) = \frac{8 \pi}{3} \rho,
\label{FE}
\end{equation}
along with matter conservation which implies $\rho a^3= \rho_{o}$ is a constant in a matter dominated universe.

To solve the equation, we specify initial conditions at the current instant of time, $a=1$, $\dot{a}=\h_{o}$, the Hubble constant and  $\overset{..}{a}= - q_{o}\h_{o}^2 $, where $q_{o}$ is the current deceleration parameter. These conditions can be used to calculate $\rho_{o}$, the current matter density (including dark matter).  Taking $q_{o}=-0.5$, which is consistent with the current observations \cite{acc_history}, the solution to eq. \ref{FE}  is plotted in figure \ref{FRW}. From the dashed-dot curve (scale factor), note that the universe started expanding at $t= -0.518 \,\h_{o}^{-1}$, from a size of $a=2/3$, prior to which it was in a contracting phase.  This is in contrast to the solution of the usual Friedmann equations which yields a big bang ($a=0$) at  $t= -0.667 \,\h_{o}^{-1}$. 
From the solid curve, note that the acceleration is currently positive but decreasing and would become negative beyond 
$t=0.319 \,\h_{o}^{-1}$. This is qualitatively consistent with the analysis of observed data in \cite{Shafieloo}, but is in sharp contrast with the standard $\Lambda$CDM model which predicts that the acceleration would continue to increase for ever. Clearly, more detailed studies are needed to understand the full implications of this formalism on cosmology.
   
In conclusion, the formalism in this paper provides a general mathematical result pertaining to affine connection and Ricci tensor in Einstein-Cartan theory in higher dimensions. It provides an alternative way to confine gravity from (D+1) to D dimensions, with modifications that can be significant to cosmology.

\textbf{Acknowledgements:} This work was supported in part (K. C. Wali) by the U. S. Department of Energy (DOE) under contract no. DE-FG02-85ER40237.

\end{document}